\documentclass[aps,prb,twocolumn,showpacs,showkeys,preprintnumbers,superscriptaddress,amsmath,amssymb,]{revtex4-1}

\usepackage{graphicx}
\usepackage{dcolumn}
\usepackage{bm}
\usepackage{hyperref}
\usepackage[dvips]{color}

\begin{document}




\title{The granularity effect in amorphous InGaZnO$_4$ films prepared by rf sputtering method}


\author{Hui Zhang}
\affiliation{Tianjin Key Laboratory of Low Dimensional Materials Physics and
Preparing Technology, Department of Physics, Tianjin University, Tianjin 300072,
China}
\author {Xin-Jian Xie}
\affiliation{School of Material Science and Engineering, Hebei University of Technology, Tianjin 300130, China}
\author{Xing-Hua Zhang}
\affiliation{School of Material Science and Engineering, Hebei University of Technology, Tianjin 300130, China}
\author{Xin-Dian Liu}
\affiliation{Tianjin Key Laboratory of Low Dimensional Materials Physics and
Preparing Technology, Department of Physics, Tianjin University, Tianjin 300072,
China}

\author{Zhi-Qing Li}
\email[Author to whom correspondence should be addressed. Electronic address: ]{zhiqingli@tju.edu.cn}
\affiliation{Tianjin Key Laboratory of Low Dimensional Materials Physics and
Preparing Technology, Department of Physics, Tianjin University, Tianjin 300072,
China}




\date{\today}

\begin{abstract}
We systematically investigated the temperature behaviors of the electrical conductivity  and Hall coefficient of two series of amorphous indium gallium zinc oxides (a-IGZO) films prepared by rf sputtering method. The two series of films are $\sim$700\,nm and $\sim$25\,nm thick, respectively. For each film, the conductivity increases with decreasing temperature from 300\,K to $T_{\rm max}$, where $T_{\rm max}$ is the temperature at which the conductivity reaches its maximum. Below $T_{\rm max}$, the conductivity decreases with decreasing temperature. Both the conductivity and Hall coefficient vary linearly with $\ln T$  at low temperature regime. The $\ln T$ behaviors of conductivity and Hall coefficient cannot be explained by the traditional electron-electron interaction theory, but can be quantitatively described by the current electron-electron theory due to the presence of granularity. Combining with the scanning electron microscopy images of the films, we propose that the boundaries between the neighboring a-IGZO particles could make the film inhomogeneous and play an important role in the electron transport processes.
\end{abstract}


\maketitle

Amorphous indium gallium zinc oxide (a-IGZO), which simultaneously possesses high optical transparency in visible range, high carrier mobility and low temperature fabrication process, is being considered as the good candidate of the active channel material in thin-film transistors (TFTs).\cite{zhang1,zhang2,zhang3,zhang4} Compared with polycrystalline metal oxides, the amorphous oxides do not have grain boundary. The grain boundary could seriously affect the carrier mobility and uniformity of oxide materials, which in turn would influence the performance of TFTs.\cite{zhang2,zhang4,zhang5,zhang6} Hence the amorphous oxide material is more suitable for TFTs than the polycrystalline one. However, even for the amorphous film, it often experiences the discontinuous to continuous process in the first  stage of film growth. Thus the boundaries between amorphous particles are also inevitable in the amorphous films. In this Letter, we measured the variations in resistivity and Hall coefficient with temperature for the thick ($\sim$700\,nm) and thin ($\sim$25\,nm) a-IGZO films from 300  down to 2 K. We found the electron-electron interaction (EEI) due to the presence of granularity governs the temperature behaviors of longitudinal and Hall transport of the a-IGZO films. Combining the scanning electron microscopy (SEM) images of the films, we conclude that the subtle influence of boundaries between the the neighboring amorphous particles in a-IGZO films on the electronic transport processes can not be neglected.

The samples were deposited on glass substrates by rf sputtering method. A commercial InGaZnO$_4$ ceramic target with purity of 99.99$\%$  was used as the sputtering source. The atomic ratio of In, Ga, and Zn is $1:1:1$. The base pressure of the sputtering chamber was below $\sim$$1\times 10^{-5}$\,Pa. The sputtering was carried out in an  argon (99.999\% in purity) atmosphere with the pressure of 0.6\,Pa, and the sputtering power was maintained at 100\,W during the deposition process. To obtain films with high conductivity, the substrate temperature was set as 688 and 788 K, respectively. We deposited IGZO films with thickness $\sim$700 and $\sim$25\,nm at each deposition temperature. The film notations are listed in Table~\ref{Table I}. Hall-bar-shaped samples were deposited for the measurements of Hall coefficient $R_H$ and conductivity $\sigma$, by using mechanical masks. The thicknesses of the films were measured with a surface profiler (Dektak, 6 M) for those films with thickness of $\sim$700\,nm, and with the low-angle x-ray diffraction for those films with thickness of $\sim$25\,nm, respectively. The crystal structures of the films were measured by using a x-ray diffractometer (XRD, D/MAX-2500v/pc, Rigaku) with Cu\,K$_\alpha$ radiation. The results indicate that all films are amorphous. The surface morphologies of the films were characterized by the SEM (S-4800, Hitachi). The electrical conductivity and Hall effect were measured using a physical property measurement system (PPMS-6000, Quantum Design), by employing the standard four-probe method. The aluminium electrodes with thickness $\sim$280\,nm were deposited before the electrical transport property measurements.

\begin{table*}
\caption{\label{Table I} Relevant parameters for the a-IGZO films. $T_{S}$ is the substrate temperature during deposition, $t$ is the mean-film thickness, $T_1$ is the maximum temperature for the $\Delta\sigma\propto\ln T$ law hold. $\sigma_{0}$ and $g_{T}$ are the adjustable parameters in Eq.~(\ref{Eq.(conductivity)}), $\sigma_{0}$ represents the conductivity without the EEI effec. $n^{\ast}$ is the measured effective carrier concentration at $T_2$,  and $T_2$ is the maximum temperature below which $R_H$ varies linearly with $\ln T$. $c_d$ and $E_{0}$ are the adjustable parameters in Eq.~(\ref{Eq.(Hall)}).}
\begin{ruledtabular}
\begin{center}
\begin{tabular}{ccccccccccccccc}
Film  &$T_{S}$   &$t$      &$T_1$    &$\sigma_{0}$       & $g_{T}$       & $T_2$   & $n^{\ast}$       &$c_d$    &$E_{0}$ \\
No.   &(K)     &(nm)      &(K)          &($10^{-4}$ S)        &                & (K)      &($10^{26}$ m$^{-3}$)     &       &($10^{-22}$ J) \\ \hline
1    &688      &653.5      &25            &5.05                &17.3            &45        &1.04                   &1.38      &7.07\\
2    &788      &683.7      &30            &4.90                 &15.5           &45        &1.00                   &1.24      &7.07\\
3    &688      &24.5       &15            &7.00                 &11.6           &40        &1.73                   &1.51      &5.07\\
4    &788      &25.6       &15            &4.42                 &7.2            &35        &1.10                   &1.29      &4.36\\
\end{tabular}
\end{center}
\end{ruledtabular}
\end{table*}

\begin{figure}[htp]
\begin{center}
\includegraphics[scale=1.0]{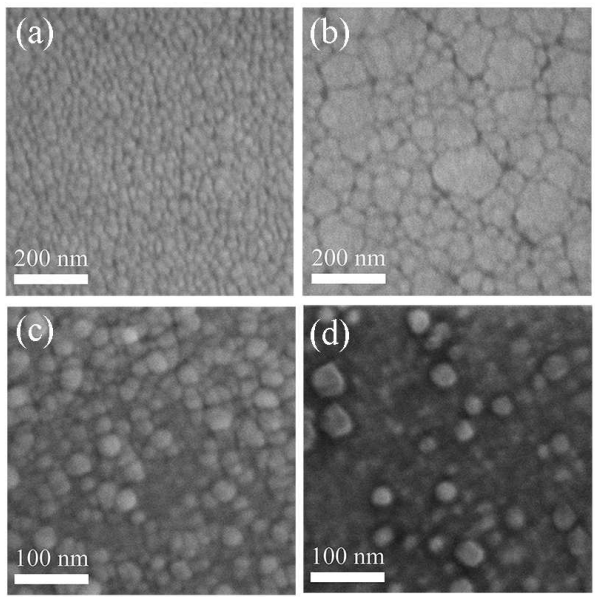}
\caption{SEM micrographs of a-IGZO films (a) No.1, (b) No.2, (c) No.3, and (d) No.4.}\label{FIGSEM}
\end{center}
\end{figure}

Figure~\ref{FIGSEM} shows the SEM images of the films. Although these IGZO films are amorphous, they are composed of amorphous IGZO particles. For those $\sim$700\,nm thick films, the mean particle size of the film deposited at 788\,K is significantly larger than that of the film deposited at 688\,K. In addition, there are distinct boundaries between the neighboring particles. These boundaries could affect the uniformity  as well as the subtle electron transport processes of the amorphous films.

\begin{figure}[htp]
\begin{center}
\includegraphics[scale=1.0]{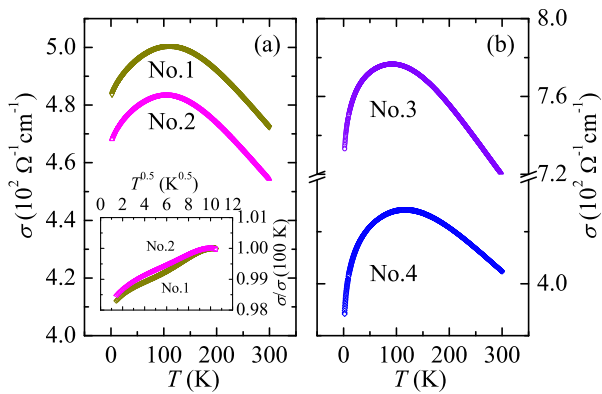}
\caption{(Color online) The conductivity as a function of temperature for a-IGZO films (a) Nos.1 and 2, and (b) Nos.3 and 4. The inset of (a): normalized conductivity $\sigma$ /$\sigma$(100 K) versus $\ln T$ under a field of 7\,T for films No.1 and No.2.}\label{3DR-T}
\end{center}
\end{figure}

Figure~\ref{3DR-T} shows the conductivity varies as a function of temperature from 300 down to 2\,K for the films. The room temperature conductivities of the films lie between $4.05\times10^{2}$ and $7.20\times10^{2}$\,$\Omega^{-1}$\,cm$^{-1}$. Hence all the films have relative high conductivity. Upon decreasing temperature from 300\,K, the conductivity of each film increases with decreasing temperature,  reaches its maximum at $T_{\rm max}$ ($T_{\rm max}$ varies from 92 to 115 K for our films), and then decreases with further decreasing temperature. The enhancement of conductivity with decreasing temperature above $T_{\rm max}$ indicates that these a-IGZO films possess metal (more strictly, degenerate semiconductor) characteristics in transport properties.\cite{zhang7,zhang8} The reasons why the a-IGZO films have the metal-like transport properties deserve further investigation. In homogeneous disordered conductors, the decrease of conductivity with decreasing temperature at liquid helium temperatures generally originates from the weak-localization (WL) and EEI effects.\cite{zhang9,zhang10,zhang11,zhang12,zhang13,zhang14} In three dimensional disordered conductors, the corrections to the conductivity due to WL and EEI effect are given as\cite{zhang12,zhang13,zhang14}
\begin{equation}\label{Eq-WL-EEI}
\Delta\sigma(T)=\frac{e^2}{2\pi^2\hbar}\frac{\alpha}{\tau_\varphi^{1/2}}+\frac{e^2}{4\pi^2\hbar}\frac{1.2}{\sqrt{2}}\left(\frac{4}{3}-\frac{3}{2}\tilde{F} \right)\frac{\sqrt{k_B T}}{\hbar D},
\end{equation}
where $\alpha$ is a constant, $e$ is the electronic charge, $\hbar$ is the Planck constant divided by $2\pi$, $k_B$ is the Boltzmann constant, $D$ is the diffusion constant, $\tau_\varphi$ is electron dephasing time, $\tilde{F}$ is the the screening factor averaged over the Fermi surface. The first term on the right hand side of Eq.~(\ref{Eq-WL-EEI}) arises from the WL effect, while the second term originates from the interaction effect. To determine $\tilde{F}$ independently, we measured the conductivity varies as a function of temperature under a magnetic field of 7\,T, which is plotted in the inset of Fig.~\ref{3DR-T}. Since the WL effect would be suppressed by the high field,\cite{zhang12,zhang14} one would expect that the conductivities vary linearly with $\sqrt{T}$ at low temperature regime. However, the $\Delta \sigma \propto \sqrt{T}$ law is not observed in the inset of Fig.~\ref{3DR-T}, which means other mechanisms should govern the temperature behavior of conductivity at liquid helium temperatures.

As mentioned above, there are distinct boundaries between the neighboring a-IGZO particles. The conductivities of the boundaries would be much less than that inside of the particles. Thus the low temperature conduction processes of the a-IGZO films would be similar to that of granular metals, where a granular metal means a metal-dielectric composite lying above the percolation threshold.\cite{zhang15,zhang16,zhang17} According to the recent theories, the EEI effect in granular metals is distinct from that of homogeneous disordered conductors. Specifically, in granular metals and the strong intergrain coupling limit [$g_{T}\gg$1, where $g_{T}=G_{T}/(2e^{2}/\hbar)$ is the dimensionless intergranular tunneling conductance, $G_{T}$ is the intergrain tunneling conductance], the conductivity\cite{zhang17,zhang18,zhang19,zhang20} and Hall coefficient\cite{zhang21,zhang22} can be written as
\begin{equation}\label{Eq.(conductivity)}
\sigma = \sigma_0 \left[ 1 - \frac{1}{2\pi g_Td} \ln \left(
\frac{g_TE_c}{k_BT} \right) \right]
\end{equation}
and
\begin{equation}\label{Eq.(Hall)}
R_H = \frac{1}{n^\ast e} \left[ 1 + \frac{c_d}{4\pi g_T} \ln \left(
\frac{E_0}{k_BT} \right) \right],
\end{equation}
where $n^{\ast}$ is the effective carrier concentration, $c_d$ is a numerical
lattice factor, $\delta$ is the mean level spacing in the metallic particle, $E_c$ is the
charging energy, $E_0$=$\min(g_TE_c, E_\text{Th})$, and $E_\text{Th}$ is the
Thouless energy, $\sigma_0$ is the conductivity without the Coulomb interaction, and $d$ is the dimensionality of the granular array. Eq.~(\ref{Eq.(conductivity)}) is valid in the temperature range $g_T\delta/k_B < T\ll E_c/k_B$, while Eq.~(\ref{Eq.(Hall)}) is valid in $g_T/k_B \delta \lesssim T \lesssim E_0/k_B$. Recently, the validity of Eq.~(\ref{Eq.(conductivity)})\cite{zhang23,zhang24,zhang25,zhang26,zhang27,zhang28,zhang29,zhang30,zhang31} and Eq.~(\ref{Eq.(Hall)})\cite{zhang28,zhang29,zhang30,zhang31} has been experimentally tested.

\begin{figure}[htp]
\begin{center}
\includegraphics[scale=1.0]{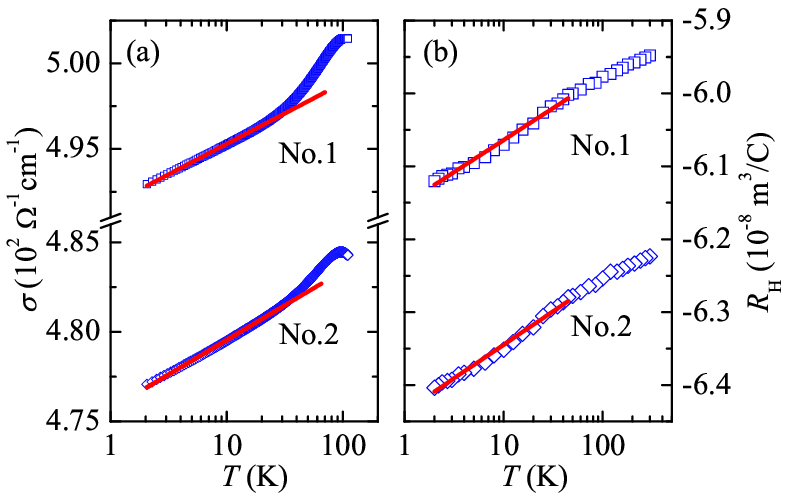}
\caption{The conductivity and Hall coefficient versus logarithm of temperature for films No.1 and No.2. The solid straight lines in (a) and (b) are least-squares fits to Eq.~(\ref{Eq.(conductivity)}) and Eq.~(\ref{Eq.(Hall)}), respectively}\label{3D-CON-Hall}
\end{center}
\end{figure}

Figure~\ref{3D-CON-Hall}~(a) and \ref{3D-CON-Hall}~(b) show the conductivities and Hall coefficients of the $\sim700$\,nm thick films variation with temperature at low temperature regime. Clearly, the conductivity (Hall coefficient) varies linearly with logarithmic temperature from 2 (2) to $\sim$25\,K ($\sim$45\,K) for each film. The experimental $\sigma$ ($R_H$) versus $T$ ($T$) data are least-squares fitted to Eq.~(\ref{Eq.(conductivity)}) [Eq.~(\ref{Eq.(Hall)})], and the results are plotted as solid curves in Fig.~\ref{3D-CON-Hall}~(a) [Fig.~\ref{3D-CON-Hall}~(b)]. When comparing the $\sigma$-$T$ data with Eq.~(\ref{Eq.(conductivity)}), $\sigma_0$ and $g_T$ are set as the adjustable parameters and the charging energy $E_c$ is taken to be $\approx$$10k_B T_1$, where $T_1\approx 25$\,K is the maximum temperature below which the $\Delta\sigma\propto\ln T$ law holds.\cite{zhang23,zhang28} For the $R_H$-$T$ fitting processes, $c_d$ and $E_0$ are the adjustable parameters, $g_T$ has been independently determined above, and $n^\ast$ is taken the value of carrier concentration at $T_2$ ($T_2$ is the maximum temperature below which  $R_H$ varies linearly with $\ln T$). The obtained fitting parameters, together with the basic parameters of sample No.1 and No.2 are listed in Table I. From Table I, one can see that the $g_T$ values are far greater than 1. Hence the necessary condition ($g_T\gg 1$) for the validity of Eqs.~(\ref{Eq.(conductivity)}) and (\ref{Eq.(Hall)}) is satisfied.\cite{zhang17,zhang23} In addition, the extracted $c_d$ values are close to 1 (more precisely, slightly greater than 1), which is consistent with the theoretical prediction.\cite{zhang23,zhang26} The obtained values of $\sigma_0$ are also reasonable. Hence Eqs.~(\ref{Eq.(conductivity)}) and (\ref{Eq.(Hall)}) are safely
applicable for films No.1 and No.2.

\begin{figure}[htp]
\begin{center}
\includegraphics[scale=1.0]{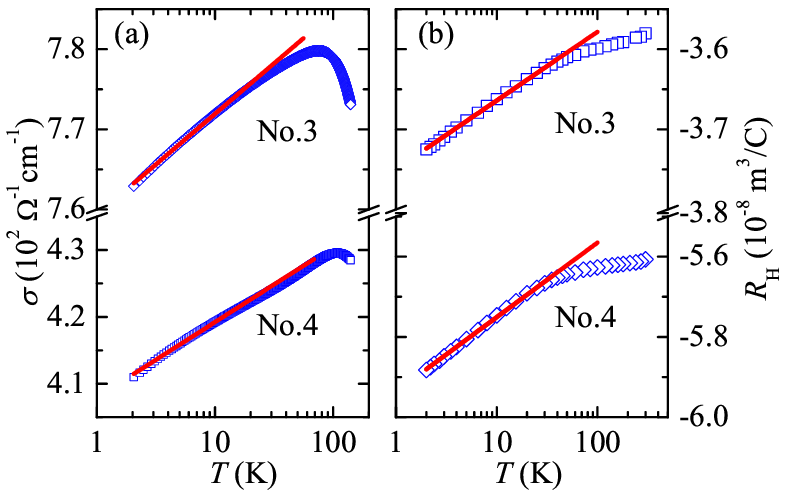}
\caption{The conductivity and Hall coefficient versus logarithm of temperature for films No.3 and No.4. The solid straight lines in (a) and (b) are least-squares fits to Eq.~(\ref{Eq.(conductivity)}) and Eq.~(\ref{Eq.(Hall)})}\label{2D-CON-Hall}
\end{center}
\end{figure}

Figure~\ref{2D-CON-Hall}~(a) and \ref{2D-CON-Hall}~(b) show variations of conductivity and Hall coefficient with temperature for the two $\sim$25\,nm, thin films, respectively. Clearly, both the conductivity and Hall coefficient vary linearly with $\ln T$ at low temperature regime. Inspection of Eqs.~(\ref{Eq.(conductivity)}) and (\ref{Eq.(Hall)}) indicates that the $\ln T$ behaviors of $\sigma$ and $R_H$ are independent of the granular array dimensionality. From the SEM images, one can see that the a-IGZO particle size lies between 20 and 30\,nm, thus it is reasonable to assume that the $\sim$25\,nm  films are covered by only one layer of a-IGZO particles. Here we treat the $\sim$25 nm films as two-dimensional (2D) granular arrays [$d=2$ in Eq.(\ref{Eq.(conductivity)})]. The experimental $\sigma(T)$ and $R_H(T)$ data are least-squares fitted to Eq.(\ref{Eq.(conductivity)}) and Eq.(\ref{Eq.(Hall)}), respectively, and the results are shown as solid curves in Fig.\ref{2D-CON-Hall}. The detailed fitting processes are identical to that used in $\sim$700\,nm films. The obtained values of parameters $\sigma_0$, $g_T$, $c_d$ and $E_0$ are listed in Table~\ref{Table I}. The values of the parameters are also reasonable.

In 2D disordered homogeneous conductors, the EEI effect can also cause a $\ln T$ correction to the conductivity. Specifically, the change of the sheet conductivity can be written as\cite{zhang14,zhang32}
\begin{equation}\label{Eq.(conductance-homogeneous)}
\Delta \sigma_{\square}(T)=\frac{e^2}{2\pi^{2}\hbar}\left(1-\frac{3}{4}\tilde{F'}\right)\ln\left(\frac{T}{T_0}\right),
\end{equation}
where $T_{0}$ is an arbitrary reference temperature, $\tilde{F'}$ is the electron screening factor. The $\sigma (T)$ data in Fig.~\ref{2D-CON-Hall} (a) are also compared with Eq.~(\ref{Eq.(conductance-homogeneous)}). Certainly, the experimental $\sigma (T)$ data follow the predication of Eq.~(\ref{Eq.(conductance-homogeneous)}). However, the least-squares fits would lead to positive $\tilde{F'}$ values for both films ($\tilde{F'}\simeq -0.17$ and $-0.13$ for films No.3 and No.4, respectively). For the 2D disordered  homogeneous conductor, the electron screening factor is required $0\leq \tilde{F'}\leq 1$.\cite{zhang9} Thus the measured $\Delta \sigma \propto \ln T$ cannot be ascribed to the conventional EEI
effect being applicable to the disordered homogeneous conductor. This result, on the other side, indicates that the boundaries between the neighboring a-IGZO particles play an important role in the electron transport processes.

In summary, we deposited $\sim$700\,nm and $\sim$25\,nm a-IGZO films by rf-sputtering methods. The SEM images indicated the films are composed of a-IGZO particles and there are distinct boundaries between neighboring  particles.  All films reveal metallic characteristics in electron transport properties. At low temperature regime, both the conductivity and Hall coefficient vary linearly with $\ln T$, which cannot be explained by the traditional EEI effect. We found that these $\ln T$ behaviors of $\sigma$ and $R_H$ result from the  EEI effect due to the presence of granularity. The subtle inhomogeneity of the a-IGZO is caused by the boundaries between neighboring amorphous particles.

This work was supported by The National Basic Research Program of China through Grant No. 2014CB931703, the NSF of China through Grant No. 11174216, and Research Fund for the Doctoral Program of Higher Education through Grant No. 20120032110065.


\begin{thebibliography}{00}\label{sec:TeXbooks}
\bibitem{zhang1}K. Nomura, H. Ohta, A. Takagi, T. Kamiya, M. Hirano, and H. Hosono, Nature \textbf{432}, 488 (2004).
\bibitem{zhang2}T. Kamiya, K. Nomura, and H. Hosono, Sci. Technol. Adv. Mater.\textbf{ 11}, 044305 (2010).
\bibitem{zhang3}H. Yabuta, M. Sano, K. Abe, T. Aiba, T. Den, H. Kumomi, and H. Hosono, Appl. Phys. Lett. \textbf{89}, 112123 (2006).
\bibitem{zhang4}T. Kamiya, K. Nomura, and H. Hosono, J. Disp. Technol. \textbf{5}, 273 (2009).
\bibitem{zhang5}C. Kim, S. Kim and C. Lee, Jpn. J. Appl. Phys. \textbf{44}, 8501 (2005).
\bibitem{zhang6}K.Nomura,H. Ohta, K. Ueda, T. Kamiya, M. Hirano, and H. Hosono, Science \textbf{300}, 1269 (2003).
\bibitem{zhang7}J. M. Ziman, Electrons and Phonons (Clarendon Press, Oxford, 1960), p.364.
\bibitem{zhang8}C. Kittel, Introduction to Solid State Physics, 7th ed, (Wiley, New York, 1996).
\bibitem{zhang9}B. L. Altshuler and A. G. Aronov, in Electron-Electron Interactions in Disordered Systems, edited by A. L. Efros and M. Pollak (Elsevier, Amsterdam, 1985).
\bibitem{zhang10}G.Bergmann, Int. J. Mod. Phys. B \textbf{24}, 2015 (2010).
\bibitem{zhang11}G.Bergmann, Phys. Rep. \textbf{107} (1984) 1.
\bibitem{zhang12}B. L. Altshuler, D. Khmelnitzkii, A. I. Larkin, and P. A. Lee, Phys. Rev. B \textbf{22}, 5142 (1980).
\bibitem{zhang13}S. P. Chiu, J. G. Lu, and J. J. Lin, Nanotechnology \textbf{24}, 245203 (2013).
\bibitem{zhang14}P. A. Lee and T. V. Ramakrishnan, Rev. Mod. Phys. \textbf{57}, 287 (1985).
\bibitem{zhang15}P. Sheng, Philos. Mag. B \textbf{65}, 357 (1992).
\bibitem{zhang16}B. Abeles et al., Adv. Phys. \textbf{24}, 407 (1975).
\bibitem{zhang17}K. B. Efetov and A. Tschersich, Europhys. Lett. \textbf{59}, 114 (2002).
\bibitem{zhang18}K. B. Efetov and A. Tschersich, Phys. Rev. B \textbf{67}, 174205 (2003).
\bibitem{zhang19}I. S. Beloborodov, K. B. Efetov, A. V. Lopatin, and V. M. Vinokur, Phys. Rev. Lett. \textbf{91}, 246801 (2003).
\bibitem{zhang20}I. S. Beloborodov, A. V. Lopatin, V. M. Vinokur, and K. B. Efetov, Rev. Mod. Phys. \textbf{79}, 469 (2007).
\bibitem{zhang21}M. Y. Kharitonov and K. B. Efetov, Phys. Rev. Lett. \textbf{99}, 056803 (2007).
\bibitem{zhang22}M. Yu. Kharitonov and K. B. Efetov, Phys. Rev. B \textbf{77}, 045116 (2008).
\bibitem{zhang23}L. Rotkina, S. Oh, J. N. Eckstein, and S. V. Rotkin, Phys. Rev. B \textbf{72}, 233407 (2005).
\bibitem{zhang24}P. Achatz, W. Gajewski, E. Bustarret, C. Marcenat, R. Piquerel, C. Chapelier, T. Dubouchet, O. A. Williams, K. Haenen, J. A. Garrido, and M. Stutzmann, Phys. Rev. B \textbf{79}, 201203 (2009).
\bibitem{zhang25}Y. C. Sun, S. S. Yeh, and J. J. Lin, Phys. Rev. B \textbf{82}, 054203 (2010).
\bibitem{zhang26}R. Sachser, F. Porrati, C. H. Schwalb, and M. Huth, Phys. Rev. Lett.  \textbf{107}, 206803 (2011).
\bibitem{zhang27}F. Porrati, R. Sachser,  C. H. Schwalb, A. S. Frangakis, and M. Huth, J. Appl. Phys. \textbf{109}, 063715 (2011).
\bibitem{zhang28}Y. J. Zhang, Z.Q. Li and J. J. Lin, Phys. Rev. B \textbf{84}, 052202 (2011).
\bibitem{zhang29}Y. Yang, Y. J. Zhang, X. D. Liu, and Z. Q. Li, Appl. Phys. Lett. \textbf{100}, 262101 (2012).
\bibitem{zhang30}Y. J. Zhang, K. H. Gao, and Z. Q. Li, Appl. Phys. Lett. \textbf{106}, 101602 (2015).
\bibitem{zhang31}Y. N. Wu, Y. F. Wei, Z. Q. Li, and J. J. Lin, Phys. Rev. B \textbf{91}, 104201 (2015).
\bibitem{zhang32}B. L. Altshuler, A. G. Aronov, and P. A. Lee, Phys. Rev. Lett. \textbf{44}, 1288 (1980).
\end{thebibliography}
\end{document}